\documentclass[10pt, conference, letterpaper]{IEEEtran}
\usepackage{cite}
\usepackage{amsmath,amssymb,amsfonts}
\usepackage{algorithmic,enumitem}
\usepackage{graphicx}
\usepackage{pdfpages}
\usepackage{textcomp}
\usepackage{xcolor}
\usepackage{multicol}
\usepackage{multirow}
\usepackage{lipsum}
\usepackage{mwe}
\usepackage{subfigure}
\usepackage{subcaption, ragged2e}
\usepackage{diagbox}
\usepackage{slashbox}
\usepackage{dblfloatfix}

\makeatletter
\newcommand*\titleheader[1]{\gdef\@titleheader{#1}}
\AtBeginDocument{%
  \let\st@red@title\@title
  \def\@title{%
    \bgroup\normalfont\large\centering\@titleheader\par\egroup
    \vskip0.6em\st@red@title}
}
\makeatother

\title{Network Traffic Analysis of Medical Devices}
\titleheader{}

\author{

\IEEEauthorblockN{Nowfel Mashnoor}
\IEEEauthorblockA{\textit{Computer Science and Engineering} \\
\textit{University of Nevada, Reno}\\
nowfel@nevada.unr.edu}
\and

\IEEEauthorblockN{Batyr Charyyev}
\IEEEauthorblockA{\textit{Computer Science and Engineering} \\
\textit{University of Nevada Reno}\\
bcharyyev@unr.edu}
}

\begin{document}

\maketitle

\begin{abstract}
The availability of medical devices such as glucose level and blood pressure measuring devices is continuously increasing. 
These devices have gained user interest as they are easy to use. 
However, medical devices introduce extra complexity to the network by being numerous, heterogeneous, and more vulnerable to cyber-attacks. 
For better network management and overall network security, it is important to understand the network traffic characteristics of the devices. 
Thus, in this paper, we analyze in detail the traffic characteristics of 8 medical devices both at the device level and at the level of individual functionality of each device.  
We collect and share both network and bluetooth traffic from a total of 51 functionalities of the devices.
Our analysis includes different metrics such as protocols, amount of incoming/outgoing traffic, DNS queries, and analysis of traffic destinations. 
We observed that devices have unique network and bluetooth traffic characteristics, that might be useful in developing networking tools for medical devices. 
\end{abstract}

\begin{IEEEkeywords}
Internet of Things, Networking, Medical Devices, Network Traffic Analysis.
\end{IEEEkeywords}

\begin{table*}[!b]
\centering
\renewcommand{\arraystretch}{1}
\caption{Medical devices used in the study.} 
\label{devicesTable}
\resizebox{0.85\textwidth}{!}{
\begin{tabular}{l|c|c|c|c}

\hline
\multirow{2}{*}{Device name} & \multicolumn{2}{c|}{Connectivity} & \multirow{2}{*}{\# of functions} & \multirow{2}{*}{Description}\\
\cline{2-3}
& Wifi & Bluetooth & \\
\hline
BabyMonitor & y & y & 7 & Real-time baby monitoring device through a video stream, manufactured by Sense-U. \\
\hline
EarWax & y & n & 5 & Device for ear wax removal with high-definition camera and LED lighting. \\
\hline
FitBitFitness & y & y & 5 & Fitness tracker that tracks daily activity, heart rate, and sleep patterns. \\
\hline
 &  &  &  & Provides 24-hour emergency support for seniors, including a wearable alert \\
GuardianAlert & y & n & 7 & button and two-way communication for immediate assistance in a crisis. \\
\hline
KardiaMobile & y & n & 6 & Device that can record a medical electrocardiogram (EKG) on the smartphone. \\
\hline
KetoScanMeter & y & y & 5 & Breath-acetone analyzer that measures the amount of ketones in a person's breath. \\
\hline
WithingsBPM & y & y & 8 & Blood pressure monitor that can be synced with a smartphone app for tracking.\\
\hline
WyzeScale & y & y & 8 & Scale that measures weight and body mass indexes. \\
\hline
\end{tabular}}
\end{table*}

\section{Introduction}
The Internet of Things (IoT) devices are continuously being integrated into our daily lives. 
These devices have various utilities that can improve quality of life through automation and monitoring. 
For instance, an IoT device \textit{ring doorbell} enables you to monitor the front door of your house or any facility, notifies you when there is a visitor through motion/sound detection, and enables you to watch the environment remotely by providing live stream from its camera. 
These devices are also different from regular devices such as laptops and mobile phones as IoT devices have more simpler nature with a limited/dedicated set of functionalities~\cite{charyyev2020iot}. 
For instance, \textit{smart blood pressure measuring device} should only have functionalities related to blood pressure measurement and provide some statistical reports about the results. 
On the other hand, IoT devices have a heterogeneous nature as there are different types of devices and numerous manufacturers following different design principles~\cite{charyyev2021modeling}.

In general, IoT devices come in the form of surveillance devices (ring doorbells, security cameras), home appliances (smart plugs, lights), and medical devices such as blood pressure measuring devices, and fitness monitoring devices. 
In particular medical devices are gaining popularity among all the people as they enable the user to monitor basic health status such as blood pressure, and body temperature without visiting a physician or doctor, and in this paper, we will mainly focus on medical devices. 
While medical devices have many use cases, cybersecurity, and privacy are the main factors that hinder its wide adoption~\cite{charyyev2021modeling}. 
Poor design principles and errors by manufacturers may lead to the production of malfunctioning devices with security vulnerabilities. 
This might be due to global competition in the IoT industry and manufacturers trying to push their devices to market as soon as possible~\cite{charyyev2021modeling}. 

Network management becomes important with the integration of medical devices into the network as these devices are more vulnerable to cyber-attacks. 
Moreover, due to the limited computational resources of IoT devices, it is challenging to install protective software on a device~\cite{sivanathan2017characterizing}. 
Thus developing networking tools tailored for medical devices enables better network management and security. 
Since network traffic-based monitoring tools such as device identification~\cite{charyyev2020locality}, and intrusion detection~\cite{hamza2019detecting} mainly use machine learning with features extracted from network data, it becomes crucial to identify protocols and characteristics of network traffic features from the devices. 
Towards this direction, it is necessary to understand the network characteristics of these devices prior to designing the dedicated networking tools and software. 
Thus in this paper, we first collect network traffic data and bluetooth data from 8 medical devices. 
Then we analyze these data on both the device level and functionality level of each device. 
Specifically, we explore how traffic data changes from device to device, and additionally how data changes for each functionality of the device. 
Our analysis includes statistical attributes such as activity cycles, port numbers, and amount of incoming/outgoing traffic data. 
We also explore application/transport layer protocols used by the devices, DNS characteristics, and contacted IP destinations. 
We believe that findings from this study will provide a better understanding of the traffic characteristics of medical devices. 
The detailed results of our analysis and data collected in our laboratory will be publicly available in our GitHub repository (\texttt{github.com/iotsec-lab/medical}). 


\vspace{-2mm}
\section{Related Works}
In this section, we provide related works on medical devices, network traffic, and bluetooth traffic analysis. 

Previous studies on medical devices focus on the adoption of medical devices in healthcare~\cite{charyyev2021modeling}, 
healthcare systems that utilize wearable electrocardiogram (ECG) monitoring devices based on bluetooth technology~\cite{kim2012mobile}, and the potential of integrating bluetooth devices and big data analysis in healthcare networks~\cite{dimitrov2016medical}.
There exist studies that focus on the ZigBee mesh protocol and enhancing the efficacy of routine monitoring procedures with medical devices~\cite{kodali2015implementation}.
There exist studies on network traffic analysis of traditional devices such as personal computers~\cite{aksoy2017operating} and mobile applications~\cite{ren2016recon}.
There also exist studies on traffic analysis of IoT devices~\cite{sivanathan2018classifying, sivanathan2017characterizing}, and drones~\cite{sciancalepore2020pinch}. 
Additional to generic traffic characterization~\cite{ren2019information, sivanathan2018classifying, sivanathan2017characterizing}, most of these studies focus on fingerprinting the network traffic for OS identification~\cite{aksoy2017operating}, device identification~\cite{charyyev2020locality}, user interaction classification~\cite{charyyev2020iot}, and intrusion detection~\cite{hamza2019detecting}. 
Studies mostly use machine learning~\cite{sivanathan2018classifying, aksoy2017operating}, natural language processing~\cite{2019Franck, 2019Enriquillo}, and hashing-based~\cite{charyyev2020locality} methods to make traffic analysis.

Existing studies on bluetooth traffic analysis focus on analyzing electrocardiogram data from wearable devices~\cite{yi2016reliability}, identification of wearable devices through bluetooth data~\cite{aksu2018identification}, 
and estimating travel time and speed on a freeway with bluetooth traffic~\cite{barcelo2010travel}. 
The methodologies to intercept bluetooth traffic are explored in~\cite{spill2007bluesniff}. 
There also exist studies focusing on privacy implications of bluetooth traffic sniffing and designing the countermeasures to sniffing~\cite{albazrqaoe2016practical}. 
Researchers also explored bluetooth based contact tracing~\cite{di2020bluetooth}. 


Compared to existing studies on medical devices~\cite{charyyev2021modeling, kim2012mobile} our work focuses on analyzing both network traffic and bluetooth traffic of medical devices. Compared to studies in network traffic analysis~\cite{aksoy2017operating, sciancalepore2020pinch, ren2019information} and bluetooth traffic analysis~\cite{yi2016reliability, aksu2018identification, sofi2016bluetooth}, our work explores how traffic characteristics vary for different functionalities of the device, and provide in-depth analysis on features and traffic destinations.

\vspace{-2mm}
\section{Methodology}
In this section, we will discuss the data collection and cleaning process. 
We collected data from 8 medical devices, and the summary of devices used in our study is provided in Table~\ref{devicesTable}. 
For each device, we first identified a set of functionalities that the device has. 
For instance, the set of functionalities for \textit{WithingsBPM} device includes turning on/off, taking blood pressure, displaying results, syncing with the server, etc. 
Then we identified if the device uses WiFi or Bluetooth connectivity. 
After that, we identified which functionalities of the device generate traffic data. 
The number of functionalities for each device that produces traffic data is provided in Table~\ref{devicesTable}. 
Then for each functionality of the device and for each possible type of connectivity (WiFi or Bluetooth), we collected data for 10 measurements. 
It is important to mention that not all device functionalities generate traffic, thus we provided only a number of functionalities that generate traffic data. Also, note that not all devices use Bluetooth connectivity.

\vspace{-2mm}
\subsection{WiFi Data}
To collect the WiFi data we connected the devices to a Linksys router flushed with OpenWrt firmware. Then we used PCAPDroid with automated Python scripts to collect the data. 
We repeated each functionality of the device 10 times and collected the data during the process. 
Some of the IoT devices generate the traffic data while they are in idle state~\cite{ren2019information}. However, we did not observe any traffic being generated when devices are not being interacted. 
Thus our network traffic is limited to the active state of the devices. 
Overall we collected data for 510 measurements that include data from 8 devices and 51 different functionalities in total. 

\begin{figure*}[!t]
\begin{center}
\hspace{-9mm}
\subfigure[Duration]{
\includegraphics[keepaspectratio=true,angle=0,width=63mm] {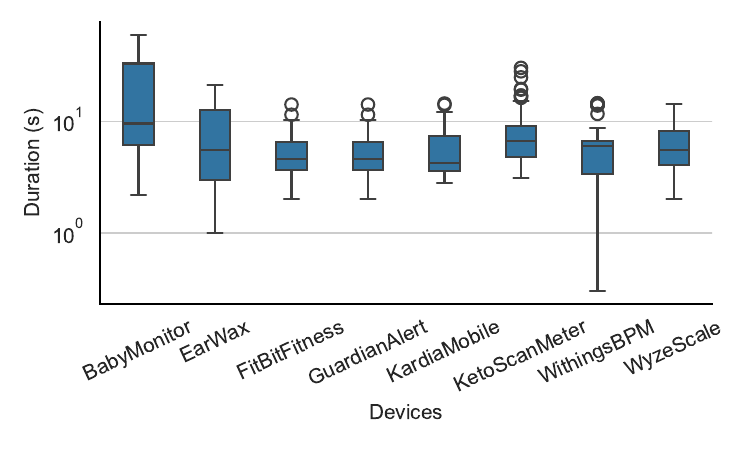}
\label{average_dur_vol_pc_1}}
\hspace{-6mm}
\subfigure[Volume]{
\includegraphics[keepaspectratio=true,angle=0,width=63mm] {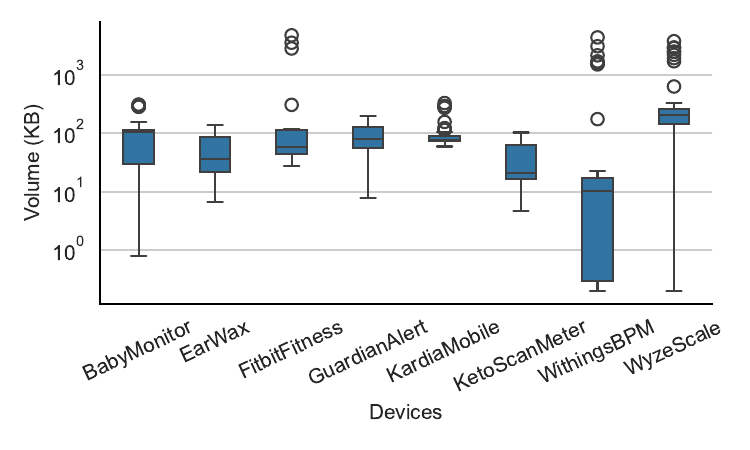}
\label{average_dur_vol_pc_2}}
\hspace{-6mm}
\subfigure[Packet Count]{
\includegraphics[keepaspectratio=true,angle=0,width=63mm] {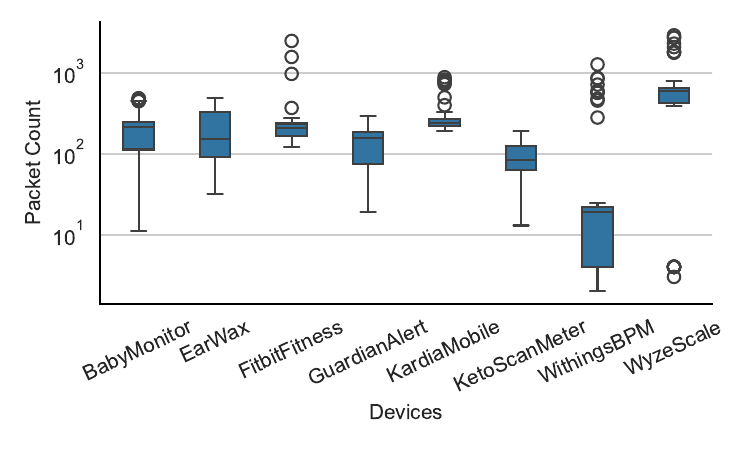}
\label{average_dur_vol_pc_3}}
\hspace{-10mm}
\vspace{-2mm}
\caption{
Average duration, volume, and packet count of traffic flows for devices, measurements include all functionalities. 
}
\label{average_dur_vol_pc}
\vspace{-7mm}
\end{center}
\end{figure*}

\begin{figure*}[!b]
\begin{center}
\hspace{-9mm}
\subfigure[Duration]{
\includegraphics[keepaspectratio=true,angle=0,width=0.5\textwidth] {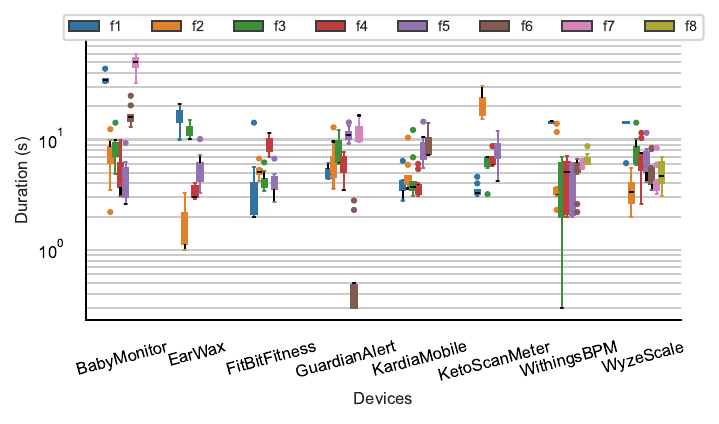}
\label{individual_dur_vol_pc_1}}
\hspace{-6mm}
\subfigure[Packet Count]{
\includegraphics[keepaspectratio=true,angle=0,width=0.5\textwidth] {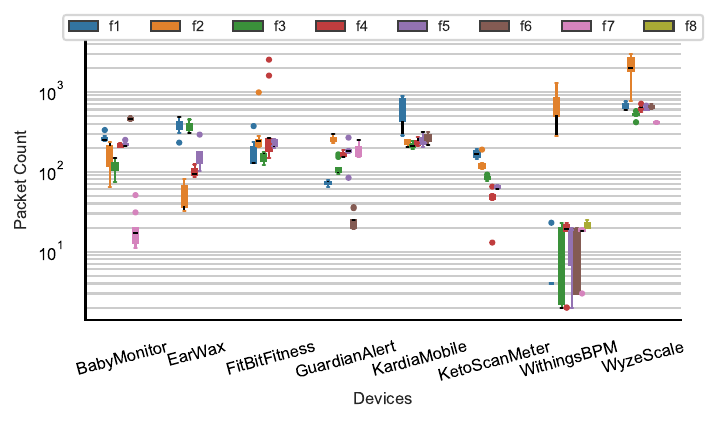}
\label{individual_dur_vol_pc_2}}
\hspace{-6mm}
\vspace{-3mm}
\caption{
Average duration and packet count for each individual functionality of the device.
}
\label{individual_dur_vol_pc}
\vspace{-5mm}
\end{center}
\end{figure*}

\vspace{-2mm}
\subsection{Bluetooth Data}~\label{blesection} 
We collected Bluetooth data from two phases of the device namely \textit{advertisement} and \textit{interaction}. 
In the advertisement phase devices send advertisement packets to connect with the base station. Once connection is established devices start to transmit data packets while the user is interacting with the medical device through a controller (mobile phone). 
In order to collect the data from medical devices, we used a nRF Sniffer~\cite{nrf}, a tool designed for intercepting Bluetooth Low Energy (BLE) packets. 
When the nRF Sniffer is used with Bluetooth Dongle connected to the data collection server, Wireshark can identify the sniffer as a separate network interface enabling us to collect the bluetooth traffic. 
First, for each device, we collected 10 advertisement data, each measurement lasting 20 seconds. 
Then, for each device, we collected 10 interaction data where we performed a set of interactions for each device in a particular order. 
Overall we collected bluetooth traffic from 100 measurements that include advertisement and interaction data from 5 medical devices.

\vspace{-2mm}
\section{Evaluation Results}
In this section, we will first present an analysis of WiFi data and then an analysis of Bluetooth data. 

\vspace{-2mm}
\subsection{Flow duration and volume}
We first evaluated the \texttt{duration}, \texttt{volume}, and \texttt{packet count} of traffic flows generated by each device using the methodology from~\cite{sivanathan2018classifying, sivanathan2017characterizing} for each metric. 
Here traffic flow is network traffic data generated by a single functionality of the device. 
The flow duration is basically the time between the first and the last packet in a flow, the volume is the total download and upload bytes, and packet count is the number of sent and received packets by the device when it is performing the functionality~\cite{sivanathan2018classifying, sivanathan2017characterizing}.
Fig.~\ref{average_dur_vol_pc} presents the average duration, volume, and packet count for all devices. 
We can see that the average duration of flows is mostly less than 10 seconds, but for some devices such as BabyMonitor and EarWax, the average duration can reach up to 59 and 21 seconds respectively.
When we look at the values for individual functionalities of the devices Fig.~\ref{individual_dur_vol_pc_1}, we see that BabyMonitor functionalities 2-5 have an average duration of less than 10s and functionality 1 (turn on) and 7 (turn off) can reach up to 60s. 
While for all devices most of the functionalities have a flow duration of less than 10s, for each device there exist certain functionalities that have a flow duration greater than 10s (Fig.~\ref{individual_dur_vol_pc_1}). 
In terms of traffic volume and packet count, we see that they follow similar patterns as expected. Overall devices exchange small amounts of data, we observed that 95\% of flows have a volume of less than 300 KB and 61\% of flows have a volume of less than 100 KB.
Interestingly, the duration and volume of the data don't show any correlation. For instance, in terms of duration, BabyMonitor has the highest average duration of the flows whereas its volume and packet count are less than WyzeScale which suggests that WyzeScale has a higher traffic rate compared to BabyMonitor. 
In terms of individual functionalities, BabyMonitor function 7 has the highest duration of the traffic flow and lowest packet count (Fig.~\ref{individual_dur_vol_pc}).
Overall it suggests that devices do not generate a large amount of data but the traffic rate can change for each device and its functionality, indicating that we can observe a high burst of network traffic for the devices.

\begin{figure*}[!b]
\begin{center}
\hspace{-9mm}
\subfigure[All devices]{
\includegraphics[keepaspectratio=true,angle=0,width=70mm] {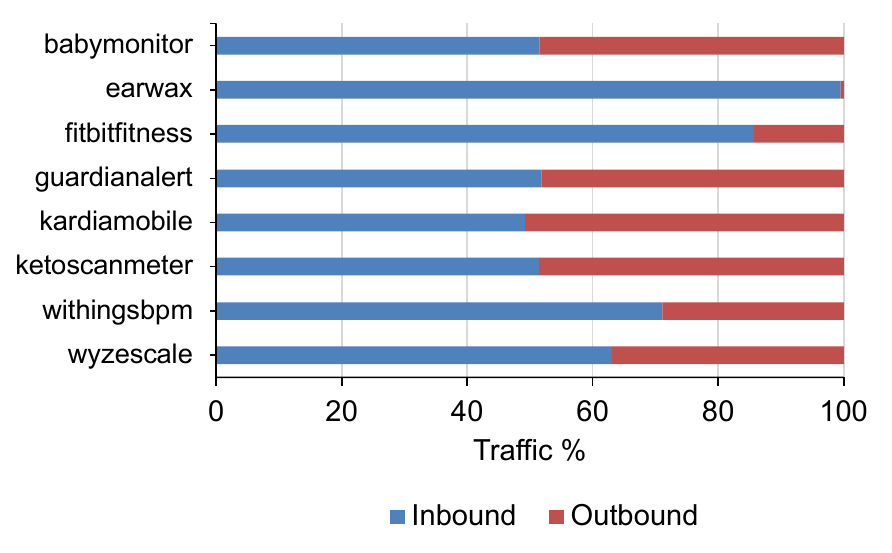}
\label{alldevice_direction}}
\hspace{-6mm}
\subfigure[BabyMonitor]{
\includegraphics[keepaspectratio=true,angle=0,width=60mm] {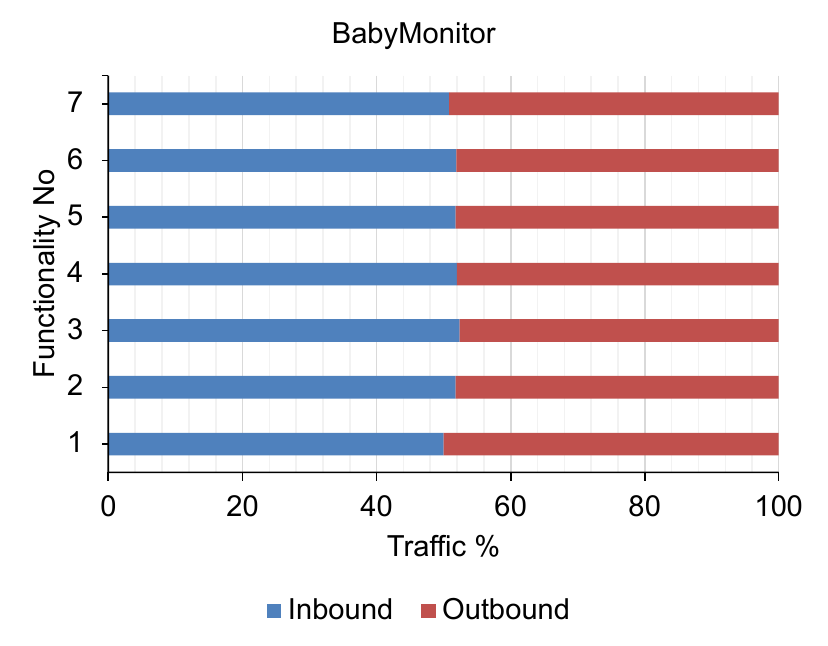}
\label{babymonitor_direction}}
\hspace{-6mm}
\subfigure[WithingsBPM]{
\includegraphics[keepaspectratio=true,angle=0,width=60mm] {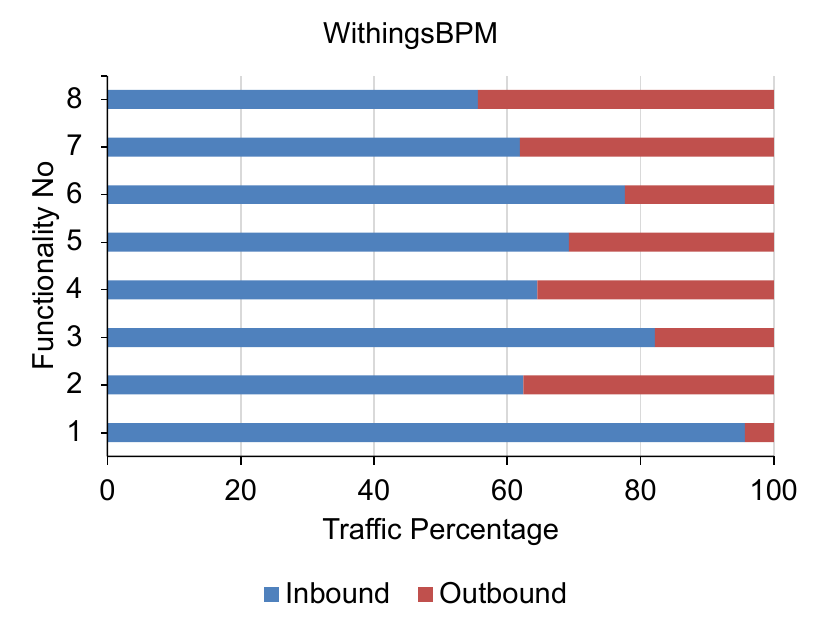}
\label{WithingsBPM_direction}}
\hspace{-10mm}
\caption{
\small Percentage of inbound and outbound network traffic for each device (a) and for each function of  BabyMonitor(b) and  WithingBPM(c).
}
\label{user_detection_for_each_method}
\vspace{-7mm}
\end{center}
\end{figure*}

\vspace{-2mm}
\subsection{Protocols and port numbers}
\begin{figure}[!t]
\includegraphics[width=0.9\columnwidth] {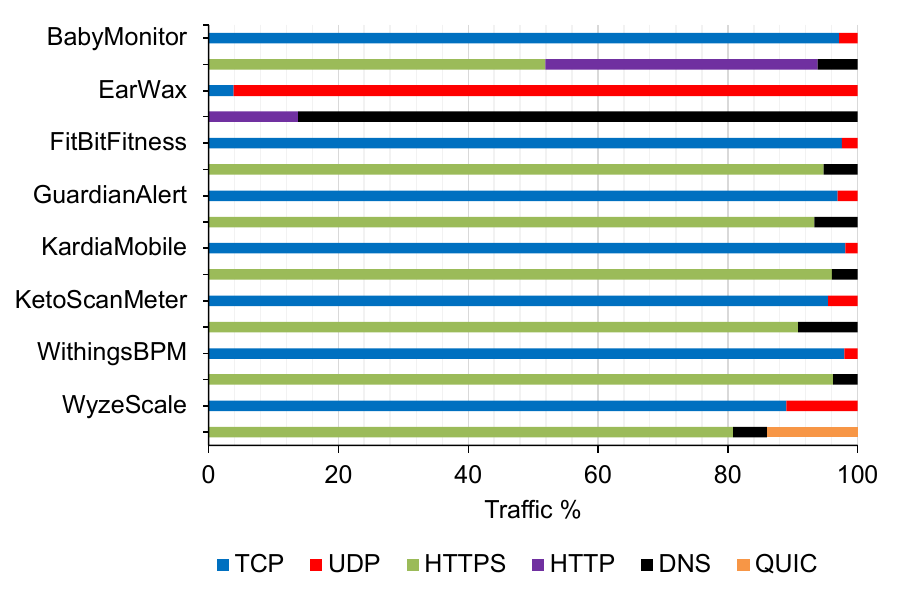}
\centering
\vspace{-2mm}
\caption{\small Protocol percentages for application and transport layer.}
\label{protocol_percent} 
\end{figure}

We analyzed the transport layer and application layer \texttt{protocols} of the devices and \texttt{port numbers} that they use.
Fig.~\ref{protocol_percent} shows for each device the percentage of transport and application layer protocols. In the figure, each device has two columns associated with it, where the first column shows the percentage of transport layer protocols and the second column shows the percentage for the application layer. 
In the transport layer, most of the devices use TCP except for EarWax and WyzeScale. 
The percentage of UDP packets in WyzeScale is around 11\% and this may correspond to DNS queries or application layer QUIC protocol that uses UDP as underlying transport layer protocol. 
For Earwax the percentage of UDP packets is higher. 
However, the operation of EarWax is slightly different compared to other devices. 
It opens its local hotspot to connect the device and mobile app then it streams the data through that channel. Thus, EarWax doesn't have much communication with the rest of the world, as we will see in the next subsection on incoming/outgoing traffic analysis. As a result, data generated by the EarWax mostly doesn't leave the local network. However, it generates some HTTP and DNS packets as an application layer protocols. 

In terms of application layer protocol, we observed that HTTPS is the dominant protocol in most of the devices. 
Every device has a certain amount of DNS communication and for WyzeScale the QUIC protocol makes around 14\% of its application layer protocol. 
Only in BabyMonitor, we observe that application layer protocol is split between HTTPS and HTTP being 51.9\% and 42.0\% of traffic flows respectively. 
In terms of destination port number, we observed that devices reach traditional port numbers such as port number 443 (used for HTTPS services over TCP) and port number 80 (used for HTTP over TCP). 
Other than traditional port we observed that BabyMonitor reaches port number 8883 in 10\% of its traffic flow and WyzeScale has a destination port number 9870 in 0.4\% of its traffic flow.
While these port numbers might be used for updates and patches they also might be used for other purposes which are not critical for device functionality.

\vspace{-2mm}
\subsection{Traffic direction}
We also analyzed the direction of the traffic generated by the devices. 
The outbound traffic is the traffic that leaves the local network and is sent to the internet. 
The inbound traffic is the traffic that is sent from the internet to the device. 
We observed that in some devices such as EarWax most of the communications are between the device and its mobile app thus data does not leave the local network. Thus for consistency, we considered those traffic as part of the inbound traffic. 
Fig.~\ref{user_detection_for_each_method} shows average inbound/outbound traffic for all devices and traffic for each functionality for the sample of devices. 
Overall we observed that for most of the devices, the traffic is inbound, meaning that communication is happening in the local network (Fig. ~\ref{alldevice_direction}). This is expected as medical devices are mostly used with their mobile app and most of the time data doesn't leave the local network except for the cases with DNS queries, firmware update checks, or fetching the data from remote servers.
For each device functionality, overall we observed that GuardianAlert, KardiaMobile, and KetoScanMeter have a similar pattern as BabyMonitor (shown in Fig.~\ref{babymonitor_direction}) where inbound and outbound traffic are split equally, and FitbitFitness shows a similar pattern to WithingsBPM (shown in Fig.~\ref{WithingsBPM_direction}) where inbound traffic is slightly higher for each device functionality.

\vspace{-2mm}
\subsection{DNS queries and traffic destinations}
\begin{figure}[!t]
\includegraphics[width=0.9\columnwidth] {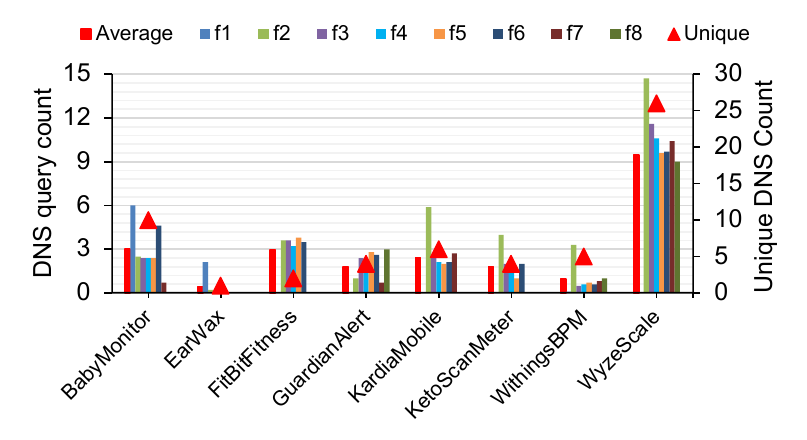}
\centering
\vspace{-4mm}
\caption{\small Average DNS query count and unique DNS count for each device and functionality.}
\label{AverageDNSQuery} 
\vspace{-6mm}
\end{figure}

Then we focused on the destination of the traffic sent to the internet and the source of the traffic coming from the internet. 
We started our analysis with DNS queries. Fig.~\ref{AverageDNSQuery} shows the average DNS query count observed across all device functionalities and for each functionality of a device. 
Figure shows EarWax is a device that has the lowest average DNS query count (0.46) and WyzeScale has the highest average DNS query count (9.45). 
We also observed that not all device functionalities make DNS queries. For instance, while BabyMonitor and WithingsBPM make DNS queries for each of their functionalities, FitBitFitness, GuardianAlert, KardiaMobile, KetoScanMeter, and WyzeScale do not make any DNS queries in their functionality 1, and only functionalities 1 and 2 of EarWax make DNS queries.
Then we analyzed the unique DNS queries for each device, results are shown as a secondary axis in Fig.~\ref{AverageDNSQuery}. 
We observed that most of the devices have 4-6 unique DNS queries, EarWax has only 1 unique DNS query which is \textit{android.bugly.qq.com} and WyzeScale has the highest number of unique DNS queries with the count of 26. 


\begin{table}[!b]
\centering
\renewcommand{\arraystretch}{1}
\caption{Destinations contacted by the devices, top domain name based on the frequency of the contact.} 
\label{DNSlabels}
\resizebox{\columnwidth}{!}{
\begin{tabular}{l|l|c|c}
\hline
Country & Organization & Contacting devices & Top Domain \\
\hline
\multirow{5}{*}{United States} & Amazon & BabyMonitor & a1.tuyaus.com \\
\cline{2-4}
 &  & GuardianAlert & ws-us2.pusher.com \\
\cline{3-4}
 &  & KardiaMobile & us-telekardia-production.alivecor.com \\
\cline{3-4}
 &  & KetoScanMeter & api.channel.io \\
\cline{3-4}
 &  & WyzeScale & wyze-platform-service.wyzecam.com \\
\cline{2-4}
 & Google & BabyMonitor & firebaselogging-pa.googleapis.com \\
\cline{2-4}
 &  & FitbitFitness & scone-pa.googleapis.com \\
\cline{3-4}
 &  & GuardianAlert & clients4.google.com \\
\cline{3-4}
 &  & KardiaMobile & api.mixpanel.com \\
\cline{3-4}
 &  & WithingsBPM & prod.rudderstack.withings.net \\
\cline{3-4}
 &  & WyzeScale & googletagmanager.com \\
\cline{2-4}
 & Alibaba & BabyMonitor & v2.sense-u.com \\
\cline{2-4}
 & Cloudflare & WyzeScale & use.fontawesome.com \\
\cline{2-4}
 & Fastly & WithingsBPM & static.withings.com \\
\cline{2-4}
 &  & WyzeScale & sdk.iad-03.braze.com \\
\cline{2-4}
 & Meta Platforms IRL & KetoScanMeter & graph.facebook.com \\
\cline{2-4}
 & Microsoft Azure & GuardianAlert & mymedicalguardian.com \\
\cline{2-4}
 & NBP NCP & KetoScanMeter & kminiapp.sentechkorea.com \\
\hline
Canada & Shopify & WyzeScale & wyzecom.myshopify.com \\
\hline
\multirow{3}{*}{China} & CNNIC & BabyMonitor & amdcopen.m.taobao.com \\
\cline{2-4}
 & China Mobile & BabyMonitor & utoken.umeng.com \\
\cline{2-4}
 & Hangzhou Alibaba Adv & BabyMonitor & offmsg.umeng.com \\
\hline
France & BSO Network Solutions & WithingsBPM & scalews.withings.net \\
\hline
\multirow{2}{*}{Hong Kong} & \multirow{2}{*}{Tencent Cloud Computing} & EarWax & \multirow{2}{*}{android.bugly.qq.com} \\
\cline{3-3}
 &  & BabyMonitor & \\
\hline
Singapore & Alibaba & BabyMonitor & ulogs.umeng.com \\
\hline
South Korea & Amazon & KetoScanMeter & ws.channel.io \\
\hline
\end{tabular}}
\end{table}

Then to find the country and organization of servers that the device is contacting, we analyzed the IP addresses returned by the DNS servers using the \texttt{WHOIS} tool.
Table-\ref{DNSlabels} shows the country and names of the organizations that devices are contacting and the top domain name for the organization.
For the organization name, we provided parent organizations such as Google for Google Cloud and Amazon for AWS.  
Overall we observe that devices reach cloud providers most of the time. 
This is expected as medical devices have limited computational power thus, they need to offload computations to remote servers. 
Cloud providers include Google, Amazon,  Microsoft, and Alibaba, and 7 devices out of 8 contacts at least Amazon or Google. 
The only device that doesn't contact these providers is EarWax which uses Tencent Cloud a cloud provider based in China. 
Even though devices are being used in the US we can see that 5 devices out of 8 reach 6 overseas countries. 
Some devices send traffic to countries where they are manufactured such as WithingsBPM to France and KetoScanMeter to South Korea.
However, BabyMonitor is made by Sense-U, a company based in California US but we can see that it sends traffic overseas to countries such as China and Hong Kong. 
Interestingly all of the organizations in China are third parties such as CNNIC (China Internet Network Information Center) a government agency in China, a telecommunication company China Mobile, and Hangzhou Alibaba Advertising.
\textbf{From a privacy perspective, this is concerning as the device being used in the US and made by a company based in the US, sends data to a government agency and telecommunication company of another country.} 
Even sending data to cloud (in particular Amazon, Google, and Microsoft) is privacy concerning as it allows them to potentially profile consumers. For example, these companies not only can learn the types of devices in a household, but also how/when they are used, simply by analyzing the traffic from devices to their cloud services~\cite{ren2019information}.

\vspace{-2mm}
\subsection{Bluetooth communication}
In addition to the WiFi connection, some medical devices (5 out of 8) also have Bluetooth connections. 
In general bluetooth devices first initiate connection through advertisement then after establishing the connection they send data packets to enable interaction. 
We collected data for both the advertisement and interaction phases and the packets are in the form of \textit{Bluetooth Low Energy Link Layer} protocol~\cite{blueWire}. 

\begin{table}[!t]
\centering
\renewcommand{\arraystretch}{1}
\caption{\small Average packet size (bytes), traffic rate (bps), and inter-arrival timing (ms) for the Bluetooth traffic of medical devices.} 
\label{BluetoothTable}
\resizebox{\columnwidth}{!}{
\begin{tabular}{l|c|c|c|c|c}

\hline
\textbf{Metric (Avg)} & \textbf{BabyMonitor} & \textbf{FitBitFitness} & \textbf{KetoScanMeter} & \textbf{WithingsBPM} & \textbf{WyzeScale} \\
\hline
\multicolumn{6}{c}{\textbf{Advertisement}} \\
\hline
packet size & 51.70 & 46.17 & 48.16 & 53.95 & 48.66 \\
\hline
inter-arrival time & 0.04 & 0.21 & 0.04 & 0.02 & 0.11 \\
\hline
traffic rate & 2345.79 & 718.90 & 1379.10 & 2358.56 & 547.21 \\
\hline
\multicolumn{6}{c}{\textbf{Interaction}} \\
\hline
packet size & 30.60 & 42.08 & 30.72 & 38.56 & 35.16 \\
\hline
inter-arrival time & 0.06 & 0.08  & 0.04 & 0.03 & 0.04 \\
\hline
traffic rate & 1177.66 & 680.53 & 816.67 & 1575.61 & 908.67 \\
\hline
\end{tabular}}
\end{table}

\begin{figure}[!b]
\includegraphics[width=0.9\columnwidth] {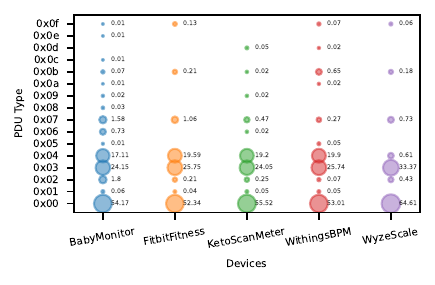}
\centering
\vspace{-4mm}
\caption{\small Percentage of observed PDU values for each device in advertisement flows.}
\label{PDUFigure} 
\end{figure}

Our goal in this subsection is to provide general characteristics of bluetooth connectivity of medical devices and explore traffic features that might be useful to better fingerprint the devices. 
Table~\ref{BluetoothTable} presents the characteristics of the bluetooth traffic in terms of the average packet size, inter-arrival timing, and traffic rate, as these features are widely used by previous studies to fingerprint the bluetooth traffic~\cite{aksu2018identification}. 
Overall we observed that bluetooth traffic is different from WiFi traffic having a lower traffic volume. 
We observed that both in the advertisement and interaction phase, BabyMonitor and WithingsBPM devices have the highest traffic rate, whereas FitBitFitness and WyzeScale have the lowest. 
Devices have unique characteristics in terms of basic traffic features such as packet size and traffic rate. 
Also, advertisement and interaction traffic flows also differ in terms of basic characteristics even if they belong to the same device. For instance, we can see that the traffic rate of BabyMonitor in the advertisement phase (2345 bps) is almost double the rate in the interaction phase (1177 bps).
Such differences in basic traffic characteristics can enable fingerprinting the device for device identification and device state identification and thus used by previous studies~\cite{aksu2018identification, yi2016reliability} for fingerprinting purposes.

Next, we explored if there exist additional features that can be used to fingerprint the medical devices through bluetooth connection. To that we extracted all 262 features listed in ~\cite{blueWire} and explored the ones that have high variance and can be used to fingerprint the devices. 
Overall we observed that most of the features have low variance or attributes corresponding to those features are left empty in the packet. 
However, there exist certain features that can be extracted from the header field of the bluetooth packet and can be used for traffic analysis. 
One of those features is \texttt{PDU type} which can be extracted from the header field of the packet. PDU type is a Protocol Data Unit that can be found in both advertisement and interaction packets. 
In general, PDU type is a 2-257 octet value and defines the type of the packet. For instance, in advertisement packets, PDU value 0000b corresponds to the Advertising Indications (ADV\_IND) packet and 0001b corresponds to the Directed Advertising (ADV\_DIRECT\_IND) packet. 
Fig.~\ref{PDUFigure} shows for each device the percentage of packets with specific PDU values. 
Overall we can see that PDU type 0x00, 0x03, and 0x04 are the most commonly observed values in all devices. We observed the same behavior in interaction flows as well. 
However, when we consider other PDU values we can see that there is a unique characteristic for each device in the amount of different PDU values. 
For instance, in advertisement flows of BabyMonitor, we can observe all PDU types except 0x0d whereas for WyzeScale we can observe only 7 out of 14 PDU types. 
Also, we observed that PDU values for BabyMonitor are different for advertisement and interaction flows (not shown in the paper), as in advertisement flows we observed 13 values out of 14, and in interaction flows we observed only 7 values out of 14. 
This suggests that in addition to classic packet size and inter-arrival timing features, the PDU types extracted from bluetooth packet header can be used to identify a device and its state. 
In addition to PDU values, we also observed certain header features such as \textit{header.flags}, \textit{header.tx\_power}, and \textit{header.llid} have high discriminatory power and can be used for various traffic analysis such as fingerprinting for network management and intrusion detection to prevent malicious activities.

\vspace{-2mm}
\section{Conclusion}
In this paper, we analyzed the network traffic characteristics of medical devices. 
We explored both wifi and bluetooth data of these devices at both device and functionality levels. 
Overall we observed that devices have unique behavior in terms of basic traffic characteristics such as volume of data, and signaling patterns. 
In terms of traffic destinations, we observed that some devices pose privacy concerns, as there are some devices contacting third parties such as advertising and overseas telecommunication companies. 
In bluetooth data, we observed that there are certain traffic features such as \textit{PDU type}, \textit{header tx power}, and \textit{header flags} that have high discriminatory power. 
Since these features can be easily extracted from packet headers they can be good alternatives and supplements to basic traffic features such as packet size and inter-arrival timing used in traffic fingerprinting and analysis.

\vspace{-2mm}
\bibliographystyle{IEEEtran}
\bibliography{references}

\end{document}